\documentstyle[12pt]{article}

\def\hybrid{\topmargin -20pt  \oddsidemargin 0pt
      \headheight 0pt   \headsep 0pt
      \textwidth 6.25in 
      \textheight 9.5in 
      \marginparwidth .875in
      \parskip 5pt plus 1pt   \jot = 1.5ex}

\hybrid

\def\x{\times}
\def\ox{\otimes}
\def\o+{\oplus}
\def\ra{\rightarrow}
\def\beq{\begin{equation}}
\def\eeq{\end{equation}}
\def\beqa{\begin{eqnarray}}
\def\eeqa{\end{eqnarray}}

\newcommand{\be}{\begin{equation}}
\newcommand{\eq}{\end{equation}}

\newcommand{\un}{\underline}

\newcommand{\LL}{{\cal L}}

\begin{document}
\thispagestyle{empty}
\rightline{IASSNS-HEP-98-28}
\rightline{hep-th/9803224}
\vspace{2truecm}
\centerline{\Large {\bf Chiral matter and transitions 
in heterotic string models}}

\vspace{1.2truecm}
\centerline{Gottfried Curio \footnote{curio@ias.edu.
The author is partially supported by NSF grant DMS 9627351}}
\vspace{.5truecm}
{\em 
\centerline{School of Natural Sciences, Institute for Advanced Study,
 Princeton, NJ 08540}}

\vspace{1.2truecm}
\vspace{.5truecm}
In the framework of $N=1$ supersymmetric string models given 
by the heterotic string on an elliptic Calabi-Yau 
$\pi :Z\ra B$ together with 
a $SU(n)$ bundle we compute the chiral matter content of the 
massless spectrum. For this purpose
the net generation number, i.e. half the third Chern class, is computed 
from data related to the heterotic vector bundle 
in the spectral cover description;
a non-technical introduction to that method is supplied. 
This invariant is, in the class of bundles considered, 
shown to be related to a discrete modulus
which is the heterotic analogue of the $F$-theory four-flux. 
We consider also the relevant
matter which is supported along certain curves in the base $B$
and derive the net generation number again from the independent 
matter-related computation. We then illustrate these considerations
with two applications.
First we show that the construction leads to numerous 
3 generation models of unbroken
gauge group $SU(5), SO(10)$ or $E_6$.
Secondly we discuss the closely related issue of the heterotic 
5-brane/instanton transition 
resp. the F-theoretic 3-brane/instanton transition.
The extra chiral matter in these transitions is 
related to the Hecke transform of the direct sum of the original bundle
and the dissolved 5-brane along the intersection of their spectral covers.
Finally we point to the corresponding $F$-theory interpretation
of chiral matter from
the intersection of 7-branes where the influence of
four-flux on the twisting along the intersection curve
plays a crucial role.
\bigskip \bigskip
\newpage

\section{Introduction}

Perhaps the most extensively studied class of $N=1$ supersymmetric
superstring compactification models is the heterotic string on a 
Calabi-Yau $Z$ with a vector bundle turned on breaking $E_8\x E_8$
to some GUT group (times a hidden $E_8$ which couples only 
gravitationally). The classical example of this construction uses
the tangent bundle, where the embedding of the spin connection in 
the gauge connection leads to an unbroken $E_6$. This was soon generalised
[\ref{W}] to the case of embedding a $SU(n)$ bundle also for $n=4$ or $5$,
leading to unbroken $SO(10)$ resp. $SU(5)$ which are also interesting 
phenomenologically. For example $E_6$ is less favoured
as its fundamental 
${\bf 27}$, decomposing as 
${\bf 16}\o+ {\bf 10}\o+ {\bf 1}$ under $SO(10)$, 
contains the ${\bf 10}\o+ {\bf 1}$ 
not observed experimentally. Similarly the $SO(10)$ is superior to the
$SU(5)$ as ${\bf 16}={\bf \bar{5}}\o+ {\bf 10}\o+ {\bf 1}$ shows that
all fermions of one generation occur in {\em one} representation 
(augmented by the right-handed neutrino). Another interesting feature 
of $SO(10)$ would be the absence
of (renormalizable) baryon number violation as $SO(10)$ forbids a 
${\bf 16}^3$ coupling.
Here we will study the heterotic string on an 
elliptically fibered Calabi-Yau $Z\ra B$ endowed with an $SU(n)$ bundle
and compute the net number of the relevant 
massless chiral matter multiplets, thereby exhibiting easily numerous 
three-generation models. As it was quite difficult to construct 
three generation models
using the tangent bundle ("embedding the spin connection in the gauge
connection"), i.e Calabi-Yau three-folds of Euler number (of
absolute value) $6$, this demonstrates impressively the greater 
flexibility of the extended ansatz.

The reason we restrict to the class of elliptic $Z$ is twofold:
first this allows one to use a fibrewise description of the bundle,
which data are then pasted together globally along the base $B$; second
one has a dual string model given by $F$-theory on a $K3$ fibered 
Calabi-Yau four-fold $X\ra B$. Both points together led to a very
satisfying description of the relevant moduli spaces and some 
non-perturbative brane-impurities to be turned on in a consistent vacuum
[\ref{FMW}],[\ref{BJPS}],[\ref{ACL}],[\ref{C}],[\ref{AC}],[\ref{CD}].

Here we will be interested in the {\em net amount of chiral matter}. 
For this
purpose we compute the net generation number (half of the third Chern 
class) of a heterotic vector bundle given in the spectral cover 
description.
As the third Chern class would vanish for $\tau$-invariant 
bundles ($\tau$ the fibre involution) the result
we get is proportional to the deviation from $\tau$-invariance as measured
by a certain discrete modulus: the $\gamma$ class, given as a
half-integer number $\lambda$ times a certain characteristic 
topological class; this is the analogue of four-flux in 
$F$-theory [\ref{CD}]. This result should be compared with an $F$-theory 
computation related to chiral matter on the intersection of two 7-branes
where, besides the twisting because of the curvedness of the compact part
of the world-volume of the 7-branes,  the influence of
4-flux on the twisting has to play a crucial role.
Note that a relation that four-flux should imply chiral matter was 
anticipated in [\ref{L}].
Actually we will also point to a second possibility to turn on $c_3(V)$:
the spectral cover construction generalizes\footnote{In [\ref{FMW}]
this did not come up as there the focus concerning Chern classes was, 
because of the consideration of 5-branes, on $c_2$ for which
essentially a consideration of a spectral cover on the 
level of surfaces $C/B$ is sufficient; 
here, as the focus is on chiral matter and so on $c_3$, we
have to face the three-fold covering $Z\x _B C \ra Z$, where then
the issue of resolution of singularities of $Z\x _B C$ occurs.} 
naturally to include a 
second discrete modulus (this time an integer $l$ which multiplies a
second universally given cohomology class, given by the exceptional
divisor resolving a naturally occuring singularity).

We also discuss the closely related framework of a 
{\em transition} where a 5-brane disappears
changing $V$ to a bundle $W$ of changed $\int _B c_2$
to restore the anomaly cancellation [\ref{FMW}]
\beqa
\int _B c_2(Z)=n_5+\sum _{i=1}^2 \int _B c_2(V_i)\nonumber
\eeqa 
Similarly on the $F$-theory side one has the process where
a 3-brane disappears by being dissolved into a finite size instanton, 
i.e. a background gauge bundle $T$ inside
the 7-brane. This exchange is governed by [\ref{BJPS}],[\ref{DM}]
\beqa
\frac{1}{24}\int_X c_4(X)=n_3+\sum_j k_j+\frac{1}{2}G^2\nonumber
\eeqa
where the contribution from the four-flux $G=\frac{1}{2\pi}dC$ 
is the 
analogue [\ref{CD}] of the contribution $-\frac{1}{2}\pi _* \gamma ^2$
inside $\int _B c_2(V)$ in the $n_5$ formula and 
$k_j=\int _{S_j} c_2(T_j)$ are the 
instanton numbers of possible background gauge bundles $T_j$ inside
the 7-brane (of compact part of world-volume given by a surface 
$S_j\subset B_3$). 
This breaks partially the gauge group which would be given, 
without background gauge bundle, by the singularity type of the 
degeneration over the 7-brane (we will consider the compact part of the 
worldvolume of the 7-brane to be given by 'sections' like 
the 'visible' two-fold base $B$ lying embedded inside the three-fold 
type IIB base $B_3$ of $F$-theory; $B_3$ is a $P^1$ bundle ober $B$). 
Note that the unbroken gauge group, heterotically
the commutator of $V$ in $E_8$, corresponds in $F$-theory to that
piece of the gauge group (related to the singularity type of $S$) which
is left over after the breaking by $T$.
We will see that in the transition extra 
chiral matter occurs if the original 
bundle had some (cf. the examples of the interesting
non-perturbative phenomenon of {\em chirality changing transitions} 
[\ref{chir}]). 
The extra chiral matter in the 
transition is related heterotically 
to the Hecke transform of the direct sum of the original bundle
and the dissolved 5-brane along the intersection of their spectral covers,
resp. in $F$-theory to the intersection of 7-branes.

In {\em section 2} we give a non-technical description of the spectral
cover method. In {\em section 3} we discuss the net generation number
(whose direct numerical evaluation from spectral cover data 
is given in the Appendix), give an independent derivation of this number
from computing the net matter supported on a certain curve and give
many 3 generation models. {\em Section 4} treates the question of 
transition and gives some heuristic dual $F$-theory considerations. 
The {\em Appendix} provides the calculation of the third Chern class.  

\section{The spectral cover description of heterotic string models}

This section provides a non-technical introduction to the spectral
cover method intended for the non-expert.

Let $\pi :Z\ra B$ be an elliptically fibered Calabi-Yau three-fold
of section $\sigma$, $V$ a $SU(n)$ vector bundle of $c_1(V)=0$ 
over $Z$. 
The idea of the spectral cover description of $V$ is to consider it
first on an elliptic fibre and then paste together these descriptions,
using global data in the base $B$. 
More precisely the data pertaining to $Z$
will be $c_1:=c_1(B)$ and $\sigma \in H^2(Z)$, $V$ will be codified
by a class $\eta \in H^{1,1}(B)$ and a half-integral number $\lambda$.
Actually, as we will describe below, the contruction generalizes easily
to include a second discrete modulus, essentially given by an integer 
$l$.\\ \\
{\bf fibre description}

We start on an elliptic fibre $E$, given in Weierstrass representation
\beqa
y^2=4x^3-g_2x-g_3\nonumber
\eeqa
with a distinguished 
reference point $p$ given, the 'origin', i.e. the zero in the 
group law, 'infinity', i.e. the point $x=y=\infty$ in the Weierstrass 
description. $V$, assumed to be fibrewise semistable,  
decomposes on $E$ as a direct sum of line bundles of degree zero.
Such a line bundle on $E$ is associated
with a unique point on $E$. The fact that the product of the line 
bundles is trivial (as we consider an $SU(n)$ bundle) translates
to the fact that the points\footnote{The line bundles resp. the
points are determined by $V$ up to permutation, i.e. up to the action 
of the Weyl group.} sum up to zero. For such an $n$-tuple of points
there exists a
unique (up to multiplication by a complex scalar) meromorphic function $w$
vanishing to first order at the points and having a pole (at most
of $n^{th}$ order) only at $p$. 
The last condition means that $w$ is a polynomial\footnote{Note that 
$x$ resp. $y$ has a double resp. triple pole at $p$; the
displayed equation is for $n$ even, for $n$ odd the last term is 
$a_nx^{(n-3)/2}y$.} in $x$ and $y$
\beqa
w=a_0+a_2x+a_3y+a_4x^2+a_5xy+\dots + a_nx^{n/2}\nonumber
\eeqa
The double interpretation of $E$ as set of points resp. parameter space
of degree zero line bundles on itself is formalized by introducing the 
Poincare bundle ${\cal P}$, a line bundle on $E_1\x E_2$ whose restriction
to $Q\x E_2$ is the line bundle on $E_2$ associated to the point 
$Q\in E_1$ (actually one uses a symmetrized version of this).
Note for later use that the naturally given involution $\tau$ on $E$ (i.e.
$a\ra -a$ in the group law) corresponds to the dualization of the bundle
on $E$.\\ \\
{\bf global description}

We assume a global choice of reference point to be given by the existence
of a section $\sigma$. The $x,y,g_2, g_3$ now become sections of the 
$2., 3., 4., 6.$ power of a line bundle over $B$ of class $c_1$ (this
is the Calabi-Yau condition). The variation over $B$ of the $n$ points
in a fibre leads to a hypersurface $C\subset Z$, a ramified $n$-fold 
cover (the 'spectral cover') of $B$ by $\pi |_C$, 
given by an equation
$s=a_0+a_2x+a_3y+\dots +a_nx^{n/2}=0$. 
Here the pole order condition leads to $s$ being a section of 
${\cal O}(\sigma )^n$ but now one can still twist by an arbitrary
line bundle ${\cal M}$ over $B$ of $c_1({\cal M})=:\eta$ so that
$s$ actually will be supposed to be a section of
${\cal O}(\sigma )^n \ox {\cal M}$. The class $\eta\in H^{1,1}(B)$ 
is the single most important
globalization datum (topological class) of the construction.
The cohomology class of $C$ in $Z$ is now given by 
\beqa
C\simeq n\sigma + \eta\nonumber
\eeqa

Now the idea in the spectral cover description of $V$ is to 
trade in the $SU(n)$ bundle $V$ over $Z$ for a line bundle ${\cal R}$
over the $n$-fold cover $p:Z\x _B C\ra Z$: the fibre of $V$ over a point 
$z\in Z$ with the $n$ preimages $\tilde{z}_i$ 
(this describes the situation outside the ramification locus) 
will be given by the sum of fibres of ${\cal R}$
at the $\tilde{z}_i$, written globally as $V=p_*{\cal R}$. If one takes
for ${\cal R}$ the global version of ${\cal P}$ one gets indeed that
{\em fibrewise} $V=p_* {\cal P}$ as $V$ was on an $E_b$ ($b\in B$) a sum
of line bundles which corresponded to an $n$-tuple of points in $E_b$
which are collected in $C_b$ and then retransformed to line bundles via
${\cal P}$ and summed up by $p_*$.
As the twist by a line bundle $L$ over $C$ leaves the fibrewise 
isomorphism class untouched ($L$ being locally trivial along $C$)
the construction generalizes to
\beqa
V=p_*(p_C^* L \ox {\cal P})\nonumber
\eeqa 
The condition $c_1(V)=0$ translates to a fixing\footnote{$\pi _*(c_1(L))=
-\pi _* \frac{c_1(C)-c_1}{2}$}
of $c_1(L)$ in 
$H^{1,1}(C)$ up to a class in 
$ker\; \pi _*:H^{1,1}(C)\ra H^{1,1}(B)$; such a class is known to be of
the form $\gamma=\lambda (n\sigma -(\eta -nc_1))$ 
with $\lambda$ half-integral. Note that the $\tau$-invariant bundles, 
i.e. bundles with $\tau ^*V=V^*$ are characterised [\ref{FMW}] by 
$\gamma =0$.

\section{Chiral matter and the generation number for 
heterotic string models}

Concerning the
net generation number $\frac{1}{2}c_3(V)=-(h^1(Z,V)-h^1(Z,V^*))$ 
for the stable vector bundle $V$ note first
that, to avoid $\tau$-invariance of $V$ which makes $c_3(V)$ trivially 
vanishing, $c_3(V)$ has to involve $\lambda$ which measures (cf. 
the remark above) the deviation from $\tau$-invariance of $V$. Actually 
one arrives from a Grothendieck-Riemann-Roch calculation for
$V=p_*(p_C^* L \ox {\cal P})$ given in the Appendix
at the relation (between numbers) 
\beqa
\frac{1}{2}c_3(V)=\lambda \eta (\eta -n c_1)\nonumber
\eeqa
{\em \un{Remark}} Actually the construction is naturally slightly 
generalised 
leading to
a dependence on a further discrete parameter: $Z\x _B C$ will have a set
${\cal S}$ of
isolated singularities when in the base the discriminant referring to
the $Z$-direction meets the branch locus referring to the $C$ direction.
Their resolution $Y\ra Z\x _B C$ leads to the possibility to formulate 
the whole construction on $Y$ and twist there by the line bundle 
corresponding to a multiple $l\in {\bf Z}$ of the exceptional divisor $E$.
Including this twist we end up with the final formula (the number 
of singularities is $|{\cal S}|=12c_1n(2\eta -(n-1)c_1)$)
\beqa
\frac{1}{2}c_3(V)=\lambda \eta (\eta -n c_1)+
\frac{l(l-1)(2l-1)}{6}|{\cal S}|\nonumber
\eeqa
Unless stated otherwise we will restrict in the following to the sector
$l=0$.\\ \\
{\bf matter}

Now let us consider an understanding of this net generation number
more directly in terms of matter, i.e. as $-(h^1(Z,V)-h^1(Z,V^*))$. Note 
[\ref{FMW}] that by the Leray spectral sequence 
$H^1(Z,V)$ is localised\footnote{This generalizes some insights of 
[\ref {BIKMSV}],[\ref{KV}] for the case of one-dimensional $B$.} 
along the fibers $E_b$ of 
$h^i(E_b,V)\neq 0$ for $i=0$ or $1$, i.e. where one of the line bundle 
summands on $E_b$ is trivial. As this corresponds to the point
at infinity this condition will not be fulfilled generically
but only along the curve
$A=C_V\cdot \sigma \subset \sigma (B)$ of cohomology class
$\eta -nc_1$ in $B\cong \sigma (B)$. 
Now $V$ restricted to $\sigma (B)$ is given 
by $\pi _* L$ 
and\footnote{note that
$deg K_C|_A=C\cdot \sigma|_C$ and $\sigma \cdot C|_{\sigma}=deg K_B|_A$
as the canonical divisors of the two surfaces
$C$ resp. $B$, which cut out $A$, are also their divisors 
$n\sigma +\eta$ resp. $\sigma$ in the Calabi-Yau $Z$ (as they represent
their normal bundles; this is for example for $B$ the relation 
$\sigma |_{\sigma}=K_B|_{\sigma}$ of the appendix); note that thereby
$\frac{1}{2}(K_C - K_B)|_A=\frac{1}{2}K_A - K_B|_A$ 
as the normal directions of $\sigma (B)$ and $C$ add up
on their intersection curve $A$, i.e. $K_A=K_B|_A+K_C|_A$.}
$c_1(L|_A)=deg \frac{K_C-K_B}{2}|_A+deg \, \gamma |_A
=\frac{1}{2}deg \, K_A - deg \, K_B|_A + \gamma \cdot \sigma$ with
$\gamma \cdot \sigma=
-\lambda \eta \sigma$ where the $\sigma$ 
becomes in $A\subset B\cong \sigma (B)$ the $\eta -nc_1$.
Consideration of the exact sequence derived from the Leray spectral 
sequence $H^p(B, R^q \pi_* V)\Rightarrow H^{p+q}(Z,V)$
\beqa
0\ra H^1(B,R^0 \pi _* V)\ra H^1(Z,V)\ra H^0(B,R^1 \pi _* V) \ra 
H^2(B,R^0 \pi _* V)\nonumber
\eeqa
which leads because of the localization here to 
$H^1(Z,V)\cong H^0(B,R^1 \pi _* V)$ leads by taking into account relative
Serre duality $\pi _* (V^*\ox \omega _{Z/B})\cong (R^1 \pi _* V)^*$ 
with the relative dualizing sheaf $\omega _{Z/B}=\pi ^* K_B^{-1}$ to the 
reinterpretation of the fermions as
$H^1(Z,V)=H^0(A,R^1 \pi _* V\, |_A)=
H^0(A, L|_A \ox K_B|_A)$ 
and considering the appropriate $\bar{\partial}$-equation on $A$, 
the index theorem gives the net-chirality as
\beqa
-\Bigl( h^0(A,L|_A \ox K_B|_A)-h^1(A,L|_A \ox K_B|_A)\Bigr) =
-\Bigl( \frac{\chi _A}{2}+deg L|_A +deg K_B|_A\Bigr) =
\lambda \eta (\eta -nc_1)\nonumber
\eeqa

As we will see shortly a number of 3 generation models with unbroken
gauge group $SU(5), SO(10)$ or $E_6$ (times some hidden group) is 
thereby easily constructed. 
For this let us first recall the connection between the breaking 
representation of $V$ and the representation of the matter charged 
under the unbroken gauge group.
\beqa
{\bf 248}&=&({\bf 3}, {\bf 27})\o+ ({\bf \bar{3}}, {\bf \overline{27}})\o+
            ({\bf 1},{\bf 78})\o+ ({\bf 8},{\bf 1})\nonumber\\
         &=&({\bf 4}, {\bf 16})\o+ ({\bf \bar{4}}, {\bf \overline{16}})\o+
            ({\bf 6},{\bf 10})\o+({\bf 1},{\bf 45})\o+ 
            ({\bf 15},{\bf 1})\nonumber\\
         &=&({\bf 5}, {\bf 10})\o+ ({\bf \bar{5}}, {\bf \overline{10}})\o+
            ({\bf 10}, {\bf \bar{5}})\o+({\bf \overline{10}}, {\bf 5})\o+
            ({\bf 1},{\bf 24})\o+ ({\bf 24},{\bf 1})\nonumber
\eeqa
So in the case of unbroken $SO(10)$ the consideration of the fundamental
$V={\bf 4}$ still suffices
(the relevant fermions are in the ${\bf 16}$ only, one does not need 
the ${\bf 10}$ which would be coupled with the 
${\bf 6}=\Lambda ^2 {\bf 4}$), whereas in the case of unbroken
$SU(5)$ one has also to consider $\Lambda ^2 V={\bf 10}$ (to get the 
${\bf \bar{5}}$ part of the fermions ${\bf 10}\o+ {\bf \bar{5}}$; the 
${\bf 10}$ and the ${\bf \bar{5}}$ will come in the same number of 
families by anomaly considerations). Note that instead of the ${\bf 27}^3$
coupling of the $E_6$ case one has in the $SO(10)$ case especially 
the ${\bf 16}\cdot{\bf 16}\cdot {\bf 10}$ with the massless Higgs bosons 
sitting in the ${\bf 10}$ to give mass to the fermions.

Note that the $\Lambda ^2 V$-related matter is localised on a
different curve [\ref{FMW}]: as $\Lambda ^2 V$ decomposes along an $E_b$
into products of pairs of those line bundles, into which $V$ decomposes, 
the triviality condition this time means that two of the points of $C_b$
should be inverses of each other (in the additive group law, 
meaning inverse bundles), so
the localization curve will be $C\cdot \tau C$, resp. its projection to
$B$.\\ \\
{\bf three-generation models}

We now give some examples of rational bases $B$ and choices of the 
globalization datum $\eta $ describing some $SU(n)$ bundle leading
to 3 generation models of unbroken gauge group\footnote{times some 
hidden gauge group left over by a choice of a 
non-trivial bundle in the other $E_8$, necessary because of 
$\eta _1+\eta _2=12 c_1$ for anomaly cancellation; note also that only 
the generation number is considerd here, leaving untouched the 
question which of these vacua can be ruled out by other requirements
or the question of breaking the GUT-group.} 
$SU(5), SO(10)$ or $E_6$. So in the following examples $n=5,4,3$ is 
understood, as well as $l=0$; also, unless stated 
otherwise\footnote{Note that in many examples 
(of $|\lambda|=\frac{1}{2}$) one can get 
the small number of standard model generations because of the possible
{\em half}-integrality of the $\gamma$ class (cf. for this issue also
[\ref{CD}],[\ref{W4flux}] and the remark at the end of section 4).},
$\lambda=-\frac{1}{2}$. 
The entries are $\eta$'s leading to $3$ generations.
\beqa
\begin{tabular}{c||c|c|c}
base & $SU(5)$ & $SO(10)$ & $E_6$ \\ \hline \hline
$F_{2k}$ & $3(1,-2+k)$ & $3(1,-3+k)$ & (0,1) \\
$F_{2k+1}$ & $-3(1,k)$ ($\lambda=\frac{1}{2}$) & 
$3(1,-3+k)$ ($\lambda=-1$) & (0,1) \\ \hline
$dP_k$ & $l-E_1-E_2$ & $l+E_1-E_2-E_3-2E_4$ & $l-E_1$ \\
 & $2E_1-E_2-E_3$ & $2E_1-E_2-E_3$ & $2E_1-E_2-E_3$ \\
 & $E_1$ & $-l+E_1+E_2$ ($\lambda=1$) & 
$l+E_1$ ($\lambda=\frac{1}{2}$)\\ \hline
$dP_6$ & & & $c_1$ \\
$dP_7$ & & $c_1$ &   \\
$dP_8$ & $-c_1$ ($\lambda=\frac{1}{2}$) &  $c_1$ ($\lambda=-1$)& \\ \hline
$dP_9$ & $b$ & $b+f$ ($\lambda=-1$) & $2b+f$ \\
       &$2b+2f$  & $-b$ ($\lambda=1$) & $-3b+f$ \\
       & $-2b$ ($\lambda=\frac{1}{2}$)  & $b+4f$ ($\lambda=1$) & $-b-2f$ 
($\lambda=\frac{1}{2}$)\\
\end{tabular}
\nonumber
\eeqa 
Here the Hirzebruch surfaces\footnote{Note that the $E_6$-models 
over the $F_m$ are derived from $6D$ models which are adiabatically
extended over a further $P^1_{(1)}$.} $F_m$ are $P^1_{(2)}$ fibrations 
over 
$P^1_{(1)}$ possesing a section of self-intersection $-m$ and have
$c_1=(2,2+m)$; above $k\geq 0$.
The $dP_k$ ($k=1, \dots , 8$) denote the del Pezzo surfaces
($P^2$ blown up in $k$ points, leading to corresponding exceptional
divisors $E_i$; they have $c_1=3l-\sum E_i$), 
and\footnote{The corresponding $Z$ is the double elliptic $CY^{19,19}$
which is a $(K3\x T^2)/{\bf Z}_2$, studied especially for example in 
[\ref{CL}]. The other Calabi-Yau three-folds over $dP_k$ resp. $F_m$
are studied in [\ref{ACL}] resp. [\ref{MV}].} 
$dP_9=\frac{1}{2}K3$ the plane 
blown up in the nine intersection points of two cubics leading
to an elliptically fibered surface with\footnote{take $b=l-E_1-E_2$, 
i.e. the proper transform of a line through two of the points} 
embedded $P^1$ base $b$ and
fibre $f=c_1$.

\section{Chiral matter and transitions in heterotic and F-theory}

We will consider now transitions where a heterotic 5-brane 
resp. a F-theory 3-brane is dissolved into an 'enlarging' of 
the heterotic bundle resp. into an instanton 
(background gauge bundle inside the 7-brane) in $F$-theory, 
leading in both descriptions to a breaking of the gauge group
to a smaller one. 
We will see that if the original bundle had some chiral matter
then in the transition a change of net-chirality occurs: namely extra
chiral matter of amount proportional to the original net-chirality
will occur.
Mathematically the computation which matches the
newly occuring chiral matter will be very similar to the discussion 
already given, the emphasis of the physical interpretation will be 
slightly different. First 
let us recall some general background pertaining to this question
[\ref{FMW}],[\ref{BJPS}]. \\ \\
{\bf the non-triviality of bundle degenerations}

One sees already from the occurence of $n$ in the
formulas for the Chern numbers that $V_n$ cannot degenerate (in the class
of bundles under consideration) to ${\cal O}\o+V_{n-1}$ (thereby
'liberating' a greater unbroken gauge group). The defining equation 
$s=a_0+a_2x+a_3y+\dots +a_nx^{n/2}=0$
for the spectral cover $C_n$ related to $V_n$ (say $n$ even) develops,
when one sends $a_n \ra 0$ to reduce to a $V_{n-1}$ situation,
besides a $C_{n-1}$ component also a second branch $\sigma$ 
(corresponding to ${\cal O}$) which {\em intersects} $C_{n-1}$. This 
means that one gets 
instead of the direct sum bundle ${\cal O}\o+ V_{n-1}$
a non-trivial extension ('elementary modification'/Hecke transform of 
${\cal O}\o+ V_{n-1}$).
By contrast $V_n$ could be possibly deformed to 
${\cal O}^{'}\o+ V_{n-1}$
with ${\cal O}^{'}$ not the trivial line bundle but a rank one sheaf
which is torsion-free, i.e. the ideal sheaf of a codim 2 subvariety:
the fibers wrapped by the heterotic 5-branes ([\ref{FMW}, 
cf. also [\ref{Sh}]).\\ \\
{\bf the connection between the line bundle
$L$ on $C$ and the gauge fields inside the $F$-theory 7-branes}

The guiding principle is that multiple components of the spectral 
cover correspond to 7-branes carrying non-abelian gauge groups inside 
them (breaking the gauge group which would otherwise correspond to the 
singularity type of the 7-brane)[\ref{BJPS}].
For this assume that $C^{'}$ is a component of $C$ of multiplicity $m >1$.
The non-reduced surface $mC^{'}$ will in general be equipped with a rank 1
sheaf $L^{'}$, one possible version of which consists in a rank $m$ 
vector bundle $M$ on the reduced surface $C^{'}$.
For example, if one has a bundle $\pi^* M$ pulled back from a bundle 
$M$ of rank $m$ on the base $B$, 
then $\pi^* M$ is fibrewise trivial of rank $m$, 
i.e. the corresponding spectral surface is $m\sigma$ 
(with spectral bundle $M$).
Now the bundle $M$ on the spectral surface
should correspond to the gauge bundle inside the $F$-theory 7-brane. 
This can also be considered from the perspective of the symmetry breaking 
mechanism [\ref{BJPS}]:
if one starts, say,
from a bundle $V=\oplus _{i=1}^n U=U\ox I_n$ (with $U$ irreducible
and $I_n$ trivial of rank $n$) then $C_V=nC_U$ and $V$ has $SU(n)$ as
an automorphism group leading to a corresponding unbroken gauge group;
then deformations of the bundle $M$ over $nC_U$ break the $U\ox I_n$
product structure, reducing thereby the automorphism group. Thus the 
bundle $M$ on a spectral surface provides the analogue to the symmetry
breking mechanism coming from a background gauge bundle inside the 
7-brane in $F$-theory.\\ \\
{\bf the transition}

Now we want to follow the transition where a $F$-theory 3-brane
disappears by being dissolved into a finite size instanton, i.e. a
background gauge bundle $\tilde{M}$ inside the 7-brane [\ref{BJPS}].
Heterotically $V$ will be enhanced to a new bundle $W$, absorbing
the instanton number of the dissolved 5-brane. 
As $W$ will be a deformation (Hecke transform) of $V\o+ \pi^* M$ and
$c_2(M)$ counts the dissolved 5-branes just as $c_2(\tilde{M})$ the
dissolved 3-branes the correspondence should be 
$M=\tilde{M}$ (both bundles are over $B$). \\ \\
{\bf heterotic situation}

We have $C_W=C_V+m\sigma$ with $m=$ rank$(M)$. To 
understand in more detail the part of the massless spectrum given by 
the deformations of $W$ look at the decomposition of the deformation 
space
\beqa
H^1(Z,End(W))&=&\;\;\;\;\;\;\; H^1(Z,End(V))\o+ 
H^1(Z,End(\pi^* M))\nonumber\\
 & &\o+ H^1(Z,Hom(V,\pi^* M))\o+ H^1(Z,Hom(\pi^* M,V))\nonumber
\eeqa
in which the first two summands correspond 
to deformations of the individual summands 
whereas the latter two summands 
(of dimensions $N$ resp. $\bar{N}$, say) deform away from the direct sum
$V\o+ \pi^* M$, the last one for example giving instead non-trivial
extensions $0\ra V\ra W\ra \pi^* M\ra 0$.
These provide the moduli of all the non-trivial Hecke transforms
of $V\o+ \pi^* M$ along the intersection curve $A$ (assumed to be 
smooth) of the spectral surfaces $C_V$ resp.
$m\sigma$ of the two summands. Note that the cohomology class of 
$A=C_V\cdot \sigma$ in $B$ is $\eta -nc_1$ and that one gets
for the net-chirality the result which depends on $M$ only through 
its rank [\ref{BJPS}]
\beqa
N-\bar{N}=\frac{1}{2}\mbox{rank}(V)c_3(\pi ^* M)-\frac{1}{2}mc_3(V)=
-\frac{1}{2}mc_3(V)\nonumber
\eeqa \\
{\bf Outlook on the $F$-theory side} 

We close with a short heuristic remark on the treatment of the 
corresponding $F$-theory situation.
Here the corresponding transition-related 
part of the spectrum is seen as follows.
The first two summands correspond essentially to complex
structure moduli of the $F$-theory $CY^4$ resp. moduli of the instanton 
background (in $F$-theory) whereas the latter two summands correspond to
chiral matter multiplets $q$ and $\tilde{q}$, transforming 
in the (bi-)fundamental 
resp. anti-fundamental of $SU(m)$, supported
on an intersection curve $A\subset B$ (assumed to be smooth) 
of the 7-branes 
(with compact part of world-volume) $B$ resp. $S$, say, corresponding
to $m\sigma$ resp. $C_V$ on the heterotic side (we assume that
inside $S$ no background gauge bundle is turned on). The index theorem
gives their net-chirality as (${\cal T}$ a twisting line bundle on $A$)
[\ref{BJPS}]
\beqa
H^0(A,M\ox {\cal T})-H^1(A,M\ox {\cal T})=
m(\frac{\chi _A}{2}+deg{\cal T})\nonumber
\eeqa
To outline some further possible steps\footnote{In yet another direction 
we would like to add that everything
in the $F$-theory side interpretation proposed here 
concerns so far the heterotic sector $l=0$ only. This clearly deserves 
further study.}
, one would have to identify now
$A$ and ${\cal T}$. Concerning the first question one should
use something like the $\eta=6c_1\pm t$
connection for $E_8$ bundles [\ref{FMW}] 
where $t$ describes\footnote{This $P^1$ bundle is given as the 
projectivization of a vector bundle
${\cal O}\o+ \Theta$ with a line bundle $\Theta$ over $B$ of $c_1=t$, so
if one uses homogeneous coordinates $a,b$, which are sections of 
${\cal O}(1)$ (of $c_1=:r$; it is 
fibrewise the usual ${\cal O}(1)$ bundle) and 
${\cal O}(1)\ox \Theta$, one has $r(r+t)=0$.
Note that then $B\cdot B=r^2=-rt$ in $B^3$ is $-t=\eta -6c_1$ in $B$.} 
the $P^1$ bundle $B^3$ over $B$.
Concerning the second question one should expect a relation of the type
$deg{\cal T}=deg \frac{K_A}{2}+ G_{\gamma}\cdot \Sigma$ 
where the four-flux 
$G_{\gamma}$, corresponding\footnote{We will always think of the
four-flux in $F$-theory as a limit from $M$-theory and not go to a
IIB interpretation of it; this allows more easily to keep contact
with the four-fold $X^4$ and thereby with the heterotic side 
(cf. [\ref{CD}]).} to the heterotic $\gamma$, is part
of the data like $S$ and $X$ which translate the heterotic bundle $V$; it
'couples' to the twisting line bundle ${\cal T}$ on $A$ by its
contribution to $deg{\cal T}$ being $G_{\gamma}\cdot \Sigma$ with 
$\Sigma$ representing the section of the $K3$ fibration $X^4\ra B$. The
idea would be then to use some relation like 
$G_{\gamma}\cdot \Sigma=\gamma \cdot \sigma$ (cf. also [\ref{CD}]).\\
{\em \un{Remark}} Concerning the critical question of 
$\lambda$-dependence 
we would like to  point to a further (cf. the $3$ generation models above)
meaning of the generator value $\lambda=\frac{1}{2}$.
For $A=\eta -nc_1$ one gets coincidence of the $F$-theory
net-chirality with the heterotic one if
$deg {\cal T}=-\frac{c_3(V)}{2}-\frac{\chi (A)}{2}=-\lambda \eta A+
\frac{K_A}{2}=(-\lambda \eta +\frac{A+K_B}{2})A=deg
\Bigl( (\frac{1}{2}-\lambda)\eta+\frac{-nc_1-c_1}{2} \Bigr) |_A$.
Now $deg {\cal T}$ is $\eta$-indep. for $\lambda =1/2$ just as 
(cf. Appendix) the $n_5$-'relevant' part $\pi ^* \omega$
of $c_2(V)=\eta \sigma +\pi ^* \omega$, $\omega \in H^4(B)$.

I thank R. Blumenhagen, D. L\"ust, T. Pantev, R. Thomas and E. Witten 
for discussion.

\newpage

\appendix

{\Large {\bf Appendix}}

\vspace{.5truecm}

{\Large {\bf Computation of the generation number from spectral 
cover data}}

\vspace{.5truecm}

{\bf the set-up}

The essential conceptual ideas of the spectral cover description of
an $SU(n)$ vector bundle over an elliptic Calabi-Yau three-fold are 
described in section 2. Here we will give the technical framework
of cohomological formulas pertaining to the Chern class computation
given later; for many details on the formulas in this subparagraph 
cf. [\ref{FMW}].

{\em Notation}: 
Let $\pi :Z\ra B$ be an elliptically fibered Calabi-Yau three-fold
of section$\sigma$, $V$ a vector 
bundle of $c_1(V)=0$ and rank $n$ 
over $Z$. $\sigma$ will also denote the class in
$H^2(Z)$ represented by the divisor $\sigma (B)$. Cohomology 
classes pulled back from $B$, like $\pi ^* \eta$, will be denoted 
by their expression on $B$, like $\eta$; similarly for
$p_C^*$ and $r$ or $\gamma$. Unspecified
Chern classes, like $c_1$, will always refer to $B$. $\pi |_C$ will 
often denoted by $\pi$ in the sequel and likewise $\sigma |_C$ 
by $\sigma$. $\Delta$ will denote the diagonal in $Z\x _B Z$ restricted 
to $Z\x _B C$, $\sigma _i$ ($i=1,2$) the classes pertaining to the 
embeddings of $B$ into the 
different fibre-factors; the class $r$ (cf. below) will always
refer to the second factor, i.e. $r=n\sigma _2+\eta +c_1$.

Now let $C\subset Z$ the spectral cover 
associated with $V$ which lies in $Z$ as a hypersurface 
given by an equation
$s=a_0+a_2x+a_3y+\dots +a_nx^{n/2}=0$. 
Here $s$ is a section of ${\cal O}(\sigma )^n \ox {\cal M}$
with ${\cal M}$ a line bundle over $B$ of $c_1({\cal M})=:\eta$.
So the cohomology class of $C$ in $Z$ is given by $n\sigma + \eta$; it is 
a ramified $n$-fold cover of $B$ by $\pi |_C$ of branch divisor
$r=-(c_1(C)-c_1)=n\sigma +\eta +c_1$. 
Now the bundle $V$ is given 
in the spectral cover description as $V=p_* (p_C^*L \otimes {\cal P})$
with $L$ a line bundle over $C$.
Here ${\cal P}={\cal O}(\Delta - \sigma _1- \sigma _2)\otimes 
{\cal Q}^{-1}$ is the suitably twisted Poincare line bundle for the 
family $Z\ra B$, ${\cal Q}=\det TB$ of $c_1({\cal Q})=c_1$ 
so that $c_1({\cal P})=\Delta - \sigma _1- \sigma _2-c_1$.
Before we move on let us collect a number of useful identities 
between the classes considered so far. One has $\sigma ^2=-\sigma c_1$ 
so that $\pi _* \sigma |_C=\eta -nc_1$; further for
$\Delta$, the diagonal in $Z\x _BZ$ restricted to $Z\x _BC$, that
$p_* \Delta =n\sigma +\eta$ and $\Delta ^2=-\Delta c_1$; moreover from
$\Delta \sigma _i=\sigma _1 \sigma _2$ one finds 
$\sigma _i c_1({\cal P})=0$ and $p_* c_1({\cal P})=0$.
\begin{center}
\begin{tabular}{ccc}
$\;\; Z\x _B C$&$\buildrel p_C \over \longrightarrow$&$C$ \\
{\tiny $p$}$ \biggl \downarrow$ & &$\;\;\;\;\;\; \biggr \downarrow $ 
{\tiny $\pi |_C$} \\
$\; Z$&$\buildrel \pi \over \longrightarrow$&$B$\\
\end{tabular}
\end{center}

To understand the Chern classes of $V$ one uses  
Grothendieck-Riemann-Roch (GRR)
$p_* (e^{c_1(p_C^*L \otimes {\cal P})} td(Z\x _B C))=ch(V) td(Z)$,
from the vanishing of whose first order term
downstairs on $Z$, i.e. from the condition $c_1(V)=0$ and the
Calabi-Yau condition for $Z$, one gets
on the $C/B$ level (i.e. restrict the consideration of GRR to 
$\sigma_1(B)\x _B C=C$ where ${\cal P}$ is trivial) the condition
$\pi _* c_1(L)=-\pi _*\frac{c_1(C)-c_1}{2}$, i.e. 
$c_1(L)=\frac{r}{2}+\gamma$ with the universal
class $\gamma=\lambda (n\sigma -\eta +nc_1)$ in 
$ker\; \pi _*:H^{1,1}(C)\ra H^{1,1}(B)$ with $\lambda$ half-integral. 
Concerning $\gamma$ note that although $\pi _* \gamma=0$ one has 
nevertheless $\pi _* \gamma ^2=-\lambda ^2n\eta(\eta-nc_1)$.\\ \\

{\bf singularities}

Actually this discussion has to be slightly modified as 
$Z\times _B C$ will have a set ${\cal S}$ of 
isolated singularities (generically ordinary double points) 
lying over points in the base $B$ where 
the branch divisor $r$ of $\pi :C \ra B$, resp. its image in $B$
$\pi_* r=n(2\eta -(n-1)c_1)$, meets the discriminant $12c_1$ 
(cf. the case of an elliptic  $K3$ over $P^1$) of $\pi: Z\ra B$. 
Their number is (as detected in $B$) 
$|{\cal S}|=12c_1n(2\eta -(n-1)c_1)$.
Let $\nu :Y\ra Z\times _B C$ be the resolution of the isolated
singularity with $E$ the exceptional divisor.
For its discussion let us make a short digression.

\begin{center}
\begin{tabular}{ccc}
$\;\; Y$& &  \\ 
{\tiny $\nu$}$ \biggl \downarrow$& & \\ 
$\;\; Z\x _B C$&$\buildrel p_C \over \longrightarrow$&$C$ \\
{\tiny $p$}$ \biggl \downarrow$ & &$\;\;\;\;\;\; \biggr \downarrow $ 
{\tiny $\pi |_C$} \\
$\; Z$&$\buildrel \pi \over \longrightarrow$&$B$\\
\end{tabular}
\end{center}

For a surface 
it is well known that the blow-up of a point $p$ leads to a divisor 
$D=P^1$ of $D^2$ being $-1$ resp. $-2$ for $p$ smooth resp. an ordinary
double point.
In general on a $n$-fold $X$ for a {\em smooth} point the local model 
of the exceptional divisor is just $P^{n-1}$ with the tautological bundle 
${\cal O}(-1)$ over it, so its self-intersection is minus a 
hyperplane $P^{n-2}$ in the $P^{n-1}$, so
in the case of a 3-fold $D^3=+1$ as $(-l)^2=+1$ for $l$ the line in the 
$P^2$.
For a {\em double point}, described by a quadratic equation 
in an ambient (n+1)-fold $W$, blow up $W$ 
at the point to get a $P^n$; 
in $X$ this would lead, at a smooth point (described by a linear 
equation), to a $P^{n-1}$, 
i.e. just the usual blow-up of $X$. But for a double point one gets
a degree 2 hypersurface in $P^n$ 
(so in the surface case a conic, i.e. still a $P^1$), again of
normal bundle ${\cal O}(-1)$, which restricted to the degree 2 locus
has degree $-2$ (so in the surface case one gets  ${\cal O}(-2)$ 
on $P^1$, i.e. self-intersection $-2$). For the 3-fold case 
$D=P^1 \x P^1$ where the $P^1$'s are described linearly in $P^3$ 
and so $D|_D=(-1,-1)$. 

So each of its $|{\cal S}|$ components 
is a divisor $D=P^1\x P^1$ of triple self-intersection $D^3=(-1,-1)^2=+2$
as $D|_D=(-1,-1)$. Furthermore 
$c_2(Y)|_D=c_2(D)+c_1(D)\cdot D|_D=4+(2,2)\cdot (-1,-1)=0$,
so $c_2$ remains unchanged as in the smooth case (cf. [\ref{GH}])
whereas the componentwise correction in $c_1(Y)$ deviates from its 
value $-2D$ in the smooth case (cf. [\ref{GH}]) as now 
$c_1(Y)|_D=c_1(D)+D|_D=(2,2)+(-1,-1)=-D|_D$, 
so here the correction is $-D$.
So one has
\beqa
c_1(Y)&=&\nu^* c_1(Z\times _B C)-E\nonumber\\
c_2(Y)&=&\nu^* c_2(Z\times _B C)\nonumber
\eeqa 

Now, more precisely, one formulates the spectral cover description 
on the resolved threefold
$Y$ which offers at the same time the possibility of
a further twisting by (a multiple $l\in {\bf Z}$ of) the canonically 
given line bundle corresponding to the exceptional divisor. 
So one arrives at the description $V=p_* \nu_* \LL$ where 
$\LL=\nu^* p_C^*L \otimes \nu^* {\cal P}\otimes {\cal O}_Y(lE)$.\\ 
In the main body of the paper we actually 
will always focus on the case $l=0$;
then the relation between $\tau$-invariance 
of $V$ and vanishing of $\gamma$
(cf. [\ref{FMW}]) holds, and $c_3(V)$ will be given by the 
$\gamma$-related term alone (cf. below). If the $l$-twist is turned
on the $E$-contribution in $c_1(Y)$ becomes important.\\ \\

{\bf the computation}

Now by Grothendieck-Riemann-Roch (GRR)
\beqa
p_* \nu_* (e^{c_1(\LL)} td(Y))&=&ch(V) td(Z)\nonumber\\
 &=&n+n\frac{c_2(Z)}{12}-c_2(V)+\frac{c_3(V)}{2}\nonumber
\eeqa
From the vanishing of the first order term
of GRR downstairs on $Z$ one gets the condition 
$(p\nu)_*c_1(\LL)=-\frac{1}{2}(p\nu)_*c_1(Y)$,
which we met above on the $C/B$ level, 
giving here $c_1({\cal L})=\frac{r}{2}+\gamma +c_1({\cal P})$.
Note further that from the Weierstrass representation one gets
$c_2(Z)=c_2+12\sigma c_1+11c_1^2=c_2+12(\sigma +c_1)c_1 -c_1^2$
[\ref{FMW}].
Similarly one can derive the expressions
$c_1(Z\x _B Z)=-c_1$ and 
$c_2(Z\x _B Z)=c_2+12[(\sigma _1+c_1)+(\sigma _2+c_1)]c_1 -c_1^2$
and from adjunction $c(Z\x _B C)=\frac{c(Z\x _B Z)}{1+n\sigma _2+\eta}$
one finally finds
\beqa
c_1(Z\x _B C)&=&-(n\sigma _2+\eta +c_1)=-r \nonumber\\
c_2(Z\x _B C)&=&(n\sigma _2+\eta +c_1)(n\sigma _2+\eta )+
c_2+12(\sigma _1+\sigma _2)c_1 +23c_1^2\nonumber
\eeqa
From the collected results one computes 
\beqa
c_2(V)=\eta \sigma -\frac{n^3-n}{24}c_1^2-
\frac{n}{8}\eta (\eta -nc_1)-\frac{1}{2}\pi _* \gamma ^2=\eta \sigma -
\frac{n^3-n}{24}c_1^2+(\lambda^2-\frac{1}{4})\frac{n}{2}
\eta (\eta -nc_1)\nonumber
\eeqa
and (for $l=0$ where only the $\lambda$-related term occurs)
\beqa
\frac{1}{2}c_3(V)&=&p_* \nu_* 
\Bigl( c_1^2({\cal L})(\frac{c_1({\cal L})}{6}+
\frac{c_1(Y)}{4})+\frac{c_1({\cal L})c_1^2(Y)}{12}+(c_1({\cal L})+
\frac{c_1(Y)}{2})\frac{c_2(Y)}{12} \Bigr) \nonumber\\
 &=&p_* \Bigl( -\frac{r^2}{24}(\gamma +c_1({\cal P}))+
\frac{(\gamma +c_1({\cal P}))^2}{6}(\gamma +c_1({\cal P}))+
\frac{c_2(Z\x _B C)}{12}(\gamma +c_1({\cal P})) \Bigr) \nonumber\\
 &=&\frac{1}{6}p_* \Bigl( 3\gamma c_1^2({\cal P})\Bigr)
=\lambda \sigma \eta (\eta -n c_1)\nonumber
\eeqa
After integration over the fiber one arrives at the relation 
between numbers\footnote{Concerning the treatment of $c_3(V)$ computation
given in the recent paper [\ref{bj}], which uses the
parabolic bundle construction, note that
the discussion given there is 
{\em very restricted}: for $n$ even it is restricted 
to $\tau$-invariant bundles of $\lambda =0$, for
which actually $c_3=0$, resp. for $n$ odd to bundles of 
$\eta \equiv 0 (n)$; 
and, especially important,
by being restricted by his ansatz to a special point in moduli space
($\lambda\sim\frac{1}{2n}$)
the author there gets as coefficient for the
main term a {\em numerical} factor ($1/n$) whose extrapolation 
(as $\sim \lambda ^2 n$ instead of the actual $\sim \lambda$) 
leads to a {\em misleading conceptual interpretation} ($\pi_*\gamma ^2$) 
of the term.}
\beqa
\frac{1}{2}c_3(V)=\lambda \eta (\eta -n c_1)\nonumber
\eeqa

If one includes also the contribution coming from $c_1({\cal O}_Y(lE))$
one gets an additional term
\footnote{This contribution is even componentwise 
integral as for example 
$\frac{l(l-1)(2l-1)}{6}=\sum_{i=1}^{l-1}i^2$.}
$\frac{l(l-1)(2l-1)}{12}2|{\cal S}|$ in 
$\frac{1}{2}c_3(V)$.

\section*{References}
\begin{enumerate}

\item
\label{W}
E. Witten, {\em New issues in manifolds of $SU(3)$ holonomy}, 
Nucl. Phys. {\bf B 268} (1986) 79.

\item
\label{FMW}
R. Friedman, J. Morgan and E. Witten, {\em Vector bundles and F-theory},
Comm Math. Phys. {\bf 187} (1997) 679, hep-th/9701162.

\item
\label{BJPS}
M. Bershadsky, A. Johansen, T. Pantev and V. Sadov, 
{\it On Four-Dimensional Compactifications of F-Theory}, hep-th/9701165.

\item
\label{ACL}
B. Andreas, G. Curio and D. L\"ust, 
{\em N=1 Dual String Pairs and their Massless Spectra},
Nucl. Phys. {\bf B 507} (1997) 175, hep-th/9705174.

\item
\label{C}
G. Curio, {\em Chiral Multiplets in N=1 Dual String Pairs},
Phys. Lett.  {\bf B 409} (1997) 18, hep-th/9705197.

\item
\label{AC}
B. Andreas and G. Curio, {\em Three-Branes and Five-Branes in N=1 
Dual String Pairs}, Phys. Lett. {\bf B 417} (1998) 41, hep-th/9706093.

\item
\label{CD}
G. Curio and R. Donagi, {\em Moduli in N=1 heterotic/F-theory duality},
hep-th/9801057, to appear in Nucl. Phys. ${\bf B}$.

\item
\label{DM}
K. Dasgupta and S. Mukhi, {\em A note on low-dimensional string 
compactifications}, Phys. Lett {\bf B 398} (1997) 285, hep-th/9612188.

\item
\label{L}
W. Lerche, {\em Fayet-Iliopoulos Potentials from Four-Folds}, 
hep-th/9709146.

\item
\label{chir}
S. Kachru and E. Silverstein, {\em Chirality changing phase transitions
in 4D string vacua}, Nucl. Phys. {\bf B 504} (1997) 272, hep-th/9704185;
E. Silverstein, {\em Closing the Generation Gap}, STRINGS'97, Amsterdam,
hep-th/9709209. I. Brunner, A. Hanany, A. Karch and D. L\"ust, {\em 
Brane Dynamics and Chiral non-Chiral Transitions}, hep-th/9801017.

\item
\label{Sh}
E. Sharpe, {\em Extremal Transitions in Heterotic String Theory},
hep-th/9705210.

\item
\label{BIKMSV}
M. Bershadsky, K. Intrilligator, S. Kachru, D. Morrison, V. Sadov and
C. Vafa, {\em Geometric singularities and enhanced gauge symmetries},
Nucl. Phys. {\bf B 481} (1996) 215, hep-th/9605200.

\item
\label{KV}
S. Katz and C. Vafa, {\em Matter From Geometry}, Nucl. Phys. {\bf B 497} 
(1997) 146, hep-th/9606086.

\item
\label{CL}
G. Curio and D. L\"ust, {\em A Class of N=1 Dual String Pairs and 
its Modular Superpotential}, Int. J. Mod. Phys. {\bf A 12} (1997) 5847, 
hep-th/9703007.

\item
\label{MV}
D. Morrison and C. Vafa, {\em Compactifications of F-Theory on 
Calabi--Yau Threefolds -- I}, Nucl. Phys.  {\bf B 473} (1996) 74,
hep-th/9602114; {\em Compactifications of F-Theory on 
Calabi--Yau Threefolds -- II}, Nucl. Phys.  {\bf B 476} (1996) 437,
hep-th/9603161.

\item
\label{W4flux}
E. Witten {\em On flux quantizaion in $M$-theory and the 
effective action}, J. Geom. Phys. {\bf 22} (1997) 1, hep-th/9609122.

\item
\label{GH}
Griffiths and Harris, {\em Principles of Algebraic Geometry}, 
John Wiley \& Sons.

\item
\label{bj}
B. Andreas, {\em On vector bundles and chiral matter in $N=1$ heterotic
compactifications}, hep-th/9802202.

\end{enumerate}

\end{document}